\documentclass[conference]{IEEEtran}
\IEEEoverridecommandlockouts
\usepackage{cite}
\usepackage{amsmath,amssymb,amsfonts}
\usepackage{algorithmic}
\usepackage{graphicx}
\usepackage{textcomp}
\usepackage{xcolor}
\usepackage{tabularx}
\usepackage{booktabs}
\usepackage{multirow}
\usepackage{makecell}
\usepackage{subcaption}
\usepackage{stfloats}
\usepackage[table]{xcolor}
\usepackage{hyperref}
\usepackage{url}
\definecolor{mycyan}{HTML}{57B5C3}
\definecolor{mycyan1}{HTML}{DCC7A9}
\definecolor{mycyan2}{HTML}{CBA67B}

\def\BibTeX{{\rm B\kern-.05em{\sc i\kern-.025em b}\kern-.08em
    T\kern-.1667em\lower.7ex\hbox{E}\kern-.125emX}}
\begin{document}

\title{AssertFix: Empowering Automated Assertion Fix via Large Language Models\\
\author{
\IEEEauthorblockN{
Hongqin Lyu\textsuperscript{1,2},
Yunlin Du\textsuperscript{3},
Yonghao Wang\textsuperscript{1,2}, 
Zhiteng Chao\textsuperscript{1},
Tiancheng Wang\textsuperscript{1},
and
Huawei Li\textsuperscript{1,2}}
\IEEEauthorblockA{\textsuperscript{1}State Key Lab of Processors, Institute of Computing Technology, CAS, Beijing, China}
\IEEEauthorblockA{\textsuperscript{2}University of Chinese Academy of Sciences, Beijing, China}
\IEEEauthorblockA{\textsuperscript{3}School of Information and Physical Sciences, University of Newcastle, Newcastle, Australia}

\IEEEauthorblockA{lvhongqin24b@ict.ac.cn}
\IEEEauthorblockA{Yunlin.Du@uon.edu.au}
\IEEEauthorblockA{\{wangyonghao22s, chaozhiteng, wangtiancheng, lihuawei\}@ict.ac.cn}
}
% {\footnotesize \textsuperscript{*}Note: Sub-titles are not captured for https://ieeexplore.ieee.org  and
}

% \and
% \and
% \and
% \and
% \and

\maketitle

\begin{abstract}
Assertion-based verification (ABV) is critical in ensuring that register-transfer level (RTL) designs conform to their functional specifications. SystemVerilog Assertions (SVA) effectively specify design properties, but writing and maintaining them manually is challenging and error-prone. Although recent progress of assertion generation methods leveraging large language models (LLMs) have shown great potential in improving assertion quality, they typically treat assertion generation as a final step, leaving the burden of fixing of the incorrect assertions to human effects, which may significantly limits the application of these methods.

To address the above limitation, we propose an automatic assertion fix framework based on LLMs, named AssertFix. AsserFix accurately locates the RTL code related to the incorrect assertion, systematically identifies the root causes of the assertion errors, classifies the error type and finally applies dedicated fix strategies to automatically correct these errors, improving the overall quality of the generated assertions. Experimental results show that AssertFix achieves noticeable improvements in both fix rate and verification coverage across the Opencore benchmarks.%这里再问一下，应该分开写还是算个平均完事了
\end{abstract}

\begin{IEEEkeywords}
functional verification, SystemVerilog assertion, assertion fix, LLM, COT 
\end{IEEEkeywords}

\section{Introduction}
In the integrated circuit (IC) design process, functional verification is a crucial step to ensure that the design logic conforms to the specifications. As the scale and complexity of designs increase, functional verification becomes increasingly important. To verify functional correctness more efficiently and comprehensively, the practice of embedding verification intent into code using SystemVerilog Assertions (SVA) has been widely adopted in IC verification. It has been reported that \cite{b1}, more than 70\% of ASIC design projects adopt ABV and the adoption percentage is still growing year by year, because ABV improves observability and controllability of designs,
and accordingly, allows for dramatic reduction in verification time and debug efforts.

Although SVA offer significant advantages in improving verification efficiency, the traditional manual process of writing SVAs still faces many challenges. For example, writing high-quality assertions not only requires a deep understanding of design specifications but also demands accurately capturing the timing relationships between signals, which is time-consuming and prone to errors. To alleviate these issues, researchers have proposed methods using LLMs to automatically generate SVAs, such as frameworks like AssertLLM\cite{b2}, which can extract information from specifications and produce corresponding assertions.

Although LLM frameworks are capable of automatically generating SystemVerilog assertions, current verification workflows generally treat generation as the endpoint\cite{b3,b4,b5,b6}. If the initial assertions contain logical errors, timing errors, or fail to fully cover the specification, engineers typically discard them and request the model to regenerate. However, regeneration often reproduces the same mistakes. This repetitive cycle not only consumes significant engineering time but also leads to gaps in verification coverage. To break this cycle of ineffective regeneration, we draw inspiration from contract repair\cite{b7} techniques in software engineering, aiming to automatically fix incorrect assertions instead of repeatedly discarding them. 

\begin{figure}
\centering
\includegraphics[width=90mm]{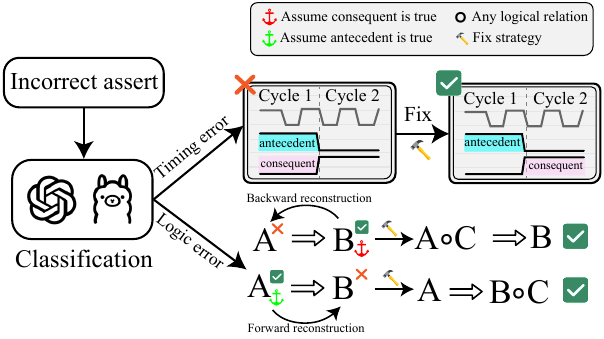}
\caption{Simplified Flow of AssertFix}
\label{Plug-and-Play Assertion Repair with AssertFix}
\end{figure}

However, transferring these techniques to the domain of SVA fixing reveals a fundamental paradigm clash. This conflict manifests on two levels: firstly, on a structural level, RTL modules run concurrently with complex interdependencies, unlike software contracts that are bound to clear function boundaries, SVAs operate on a set of discrete signals scattered across various modules. Secondly, and more critically, at the logical level, software repair techniques rely on static logic, whereas hardware logic is inherently coupled with timing constraints. Thus, this fundamental difference means that existing techniques lack the core vocabulary to understand clocks, cycle delays, or sequential implication, and as a result, they are unable to perform a root-cause diagnosis of failures.

To address the aforementioned dual challenges at both structural and logical levels, and to enable the precise fix of incorrect assertions, we propose the AssertFix framework, shown as Fig. \ref{Plug-and-Play Assertion Repair with AssertFix}. First, to tackle the problem of structural ambiguity, the framework combines Retrieval-Augmented Generation (RAG)\cite{b8} technology with prompts that guide the model's Chain-of-Thought (CoT)\cite{b9} reasoning to precisely locate the RTL logic highly relevant to the assertion. Subsequently, to resolve logical-level conflicts, AssertFix employs targeted fix strategies: for timing errors, it leverages LLM-based cycle simulation; For logic errors, it applies a bidirectional reconstruction strategy based on the Control
and Data Flow Graph (CDFG)\cite{b10}, enabling precise assertion fixing. Our contributions in this work are summarized below:
\begin{enumerate}[]
    \item To the best of our knowledge, AssertFix is the first automatic fix framework specifically targeting SVAs generated by LLMs. By identifying relevant logic code, classifying assertion errors, and applying targeted fix strategies, AssertFix achieves semantic correction of incorrect assertions.
    \item For the logic fix task, we propose a bidirectional anchor reconstruction strategy, which effectively identifies the root causes of incorrect assertions and guides the model to generate semantically consistent and logically accurate fixes.
    \item We evaluated AssertFix on four benchmark designs from OpenCores. Experimental results show that AssertFix achieves noticeable improvements in
both fix rate and verification coverage (COI and proof core) across the design.

\end{enumerate}

%贡献最后回过头来再写，先往下修，把下面都修完再写贡献也不迟
\section{related work}
\begin{figure*}
\centering
\includegraphics[width=170mm]{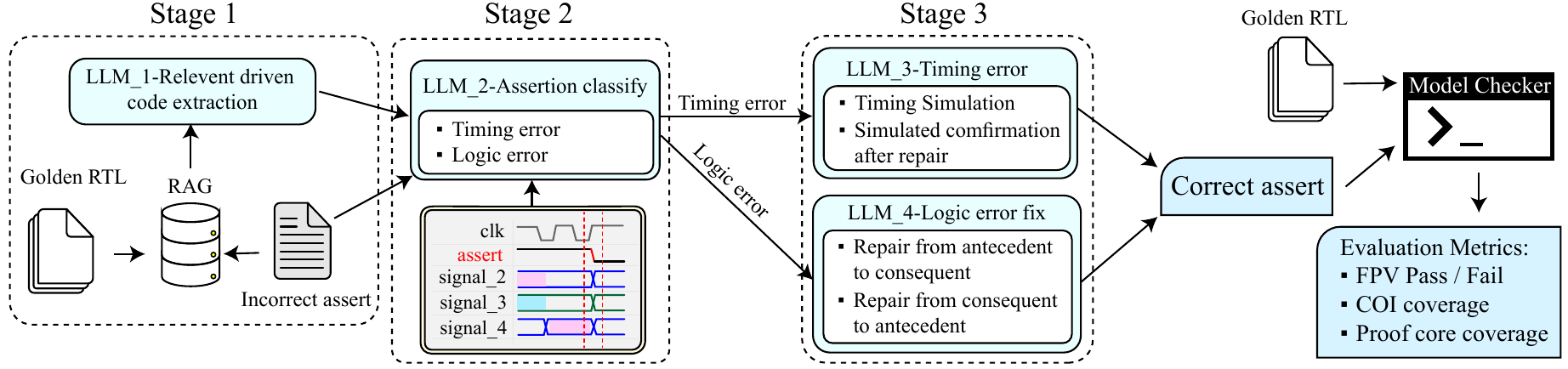}
\caption{AssertFix workflow. The fix of incorrect assertions consists of three steps: extracting relevant RTL logic using RAG, classifying assertion errors as either timing  or logic errors based on waveform and code context, and performing guided assertion fix tailored to the identified error type. }
\label{AssertFixfarmework}
\end{figure*}
Since the early days of software engineering, researchers have explored repairing specifications to address defects caused by inconsistencies between implementations and specifications. Contract repair, as one such approach, automatically adjusts program specifications to resolve conflicts between code and its specifications, rather than modifying the code itself.

%从Exploring Automatic Specification Repair in Dafny Programs 里面去找参考文献，这个还是可以的
Research in contract repair has progressed through various stages. Early work by Gopinath et al.\cite{b11} utilized Alloy specifications to repair Java programs. Following this, Wei et al.\cite{b12} introduced AutoFix, a system that automatically repairs contracts in Eiffel programs.  In the educational context, there are also works that use a reference implementation of a program as an oracle to fix students’ submissions\cite{b13}.

The scope was later extended to fixing inconsistencies in OCL constraints\cite{b14} and applying template-based or test-driven repairs for Alloy specifications\cite{b16}. These efforts demonstrated the feasibility of repairing specifications rather than code. More recently, LLM-based approaches, such as the dual-agent framework by Alhanahnah et al.\cite{b17}, have shown promising results in repairing Alloy specifications, with GPT-4 outperforming traditional tools.

% Although contract repair techniques have been extensively studied and have achieved significant results in software engineering, these methods have not yet been applied to specification or assertion repair in hardware design, especially RTL designs. Therefore, our research extends the idea of contract repair from the software domain into hardware verification for the first time, proposing an automated assertion repair method specifically for RTL designs to fill this gap and enhance the efficiency of hardware verification.%这里可以这么写吗？让我觉得有点难以抉择
\section{AssertFix Farmework}
To effectively fix incorrect assertions, we propose AssertFix, a framework that establishes a multi-agent collaborative mechanism by integrating waveforms, incorrect assertions, and RTL code. The overall architecture of AssertFix is depicted in Fig.\ref{AssertFixfarmework}. AssertFix guides LLM to classify and semantically fix incorrect assertions, ensuring semantic alignment between assertions and the intended design behaviors. Ultimately, AssertFix produces corrected assertions that are both logically accurate and behaviorally consistent. 

To effectively locate RTL code logic segments related to incorrect assertions and apply appropriate fix strategies for different error types, the assertion fix process is decomposed into three stages:
\begin{enumerate}[]
    \item \textbf{Relevance-driven Code Extraction:} Given the complexity of RTL designs, we combine a RAG mechanism with LLM to retrieve and filter RTL logic segments related to incorrect assertions.
    \item \textbf{Assertion Error Type Classification:} In this stage, we feed the incorrect assertion, its associated waveform, and the RTL logic into the LLM to classification whether the error is timing- or logic-related. 
    \item \textbf{Incorrect Assertion Fixing:} Each identified error type is mapped to a corresponding fix strategy to generate a correct assertion with consistent timing and logic. %我感觉这里的语言描述应该还是需要进行一定的修改的  这里的assert fix是需要修改的
\end{enumerate}
\subsection{Stage 1: Relevance-Driven Code Extraction}%这里根本就不是RTL的抽象，抽象是相当于修改代码的
Given the complexity and considerable length of RTL designs, accurately localizing the RTL logic relevant to an incorrect assertion is non-trivial. For example, an incorrect assertion often involves multiple antecedent and consequent signals, which may themselves be 
\begin{figure}
\centering
\includegraphics[width=0.8\linewidth]{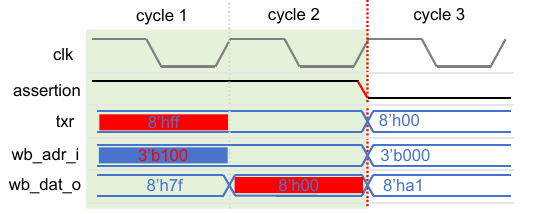}
\caption{Logic error assertion showing a timing-like waveform.}
\label{class ex}
\end{figure}
intricately connected to various other logic components in the RTL design. Without narrowing the analysis scope, directly attempting to localize the RTL logic relevant to the assertion can introduce substantial noise due to the coupling between signals. This noise interferes with the LLM’s ability to reason about the core causal path, making it difficult to accurately identify the root cause. 

To effectively extract RTL segments relevant to the incorrect assertion. This stage use a two-step filtering approach: first, a RAG mechanism combined with a query $q$ is used to coarsely retrieve RTL code segments potentially related to the incorrect assertion. Then, an LLM is employed to perform fine-grained filtering on the retrieved code segments.%这里的抽象都需要进行修改

The coarse-grained retrieval step leverages a RAG mechanism to initially identify RTL code segments relevant to the incorrect assertion. Specifically, the core principle of RAG is to select the most relevant subset from a preprocessed collection of code chunks based on relevance computations with respect to a user query $q \in \mathcal{Q}$. The query follows a uniform template “\emph{What are the code snippets related to \{signal\}?},” where the signal used in the template to start stage 1 is selected from the the consequent part of the incorrect assertion. The rationale behind this choice is that, if the core of incorrect stems from a logic-related error, tracing backward from the consequent signal to its COI will likely expose logical inconsistencies with respect to the assertion condition.

After using RAG to perform coarse-grained retrieval on the RTL code, we further using a COT prompt to guide the fine-grained filtering process. Specifically, we first prune obviously irrelevant signals and declarations to significantly reduce contextual complexity, allowing the model to efficiently focus on impactful RTL logic. Next, we perform a semantic-level logical consistency check to remove code segments that are inconsistent with the assertion semantics, eliminating potential logical distractions and enabling the model to focus on the critical RTL logic that truly causes the assertion failure. By combining coarse and fine-grained retrieval, the model can obtain RTL segments that are consistent with the assertion semantics and related to failures.
\subsection{Stage 2: Assertion Error Type Classification}
\begin{figure}[!t]
\centering
\includegraphics[width=0.7\linewidth]{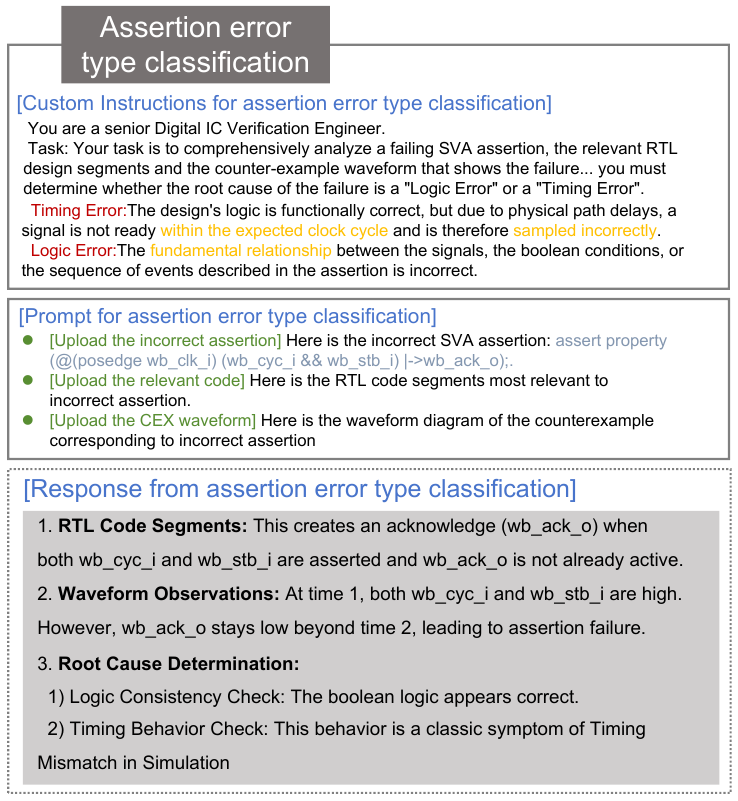}
\caption{Prompt for assertion classification}
\label{Prompt example for assertion classify}
\end{figure}
Through our observation, incorrect assertions generally fall into two major categories: timing errors and logic errors. Timing errors refer to situations where the behavior described by an assertion does not occur precisely at the expected time, such as happening one cycle earlier or later, resulting in a check failure. Logic errors, on the other hand, arise when the assertion antecedent or consequent is fundamentally inconsistent with the RTL semantics, for example, due to missing signal dependencies or incorrect logical relationships. A correct classification enables the selection of appropriate fix strategies for different error types, thereby improving the overall rate of the assertion fix process.

%下面这个第一段描述需要去看看有没有文献支撑
In practical verification workflows, classifying incorrect assertions based solely on the assertion itself and the associated counterexample waveform can lead to misclassification. This is because the waveform may exhibit misleading timing characteristics, causing genuine logic errors to be incorrectly classified as timing errors. Consider the following assertion as an example:
{\small
\begin{align*}
&(\texttt{wb\_adr\_i} == 3'b100) \rightarrow \ \#\#1 (\texttt{wb\_dat\_o} == \texttt{\$past(txr)})
\end{align*}
}
As depicted in Fig.\ref{class ex}, the counterexample waveform shows wb\_dat\_o apparently lagging one cycle behind txr. Such superficial timing delays can easily mislead the LLM into interpreting this as a timing error. However, the assertion itself contains a logical flaw: the correct antecedent condition should be wb\_adr\_i == 3'b101 instead of the current 3'b100.

To eliminate potential misclassification of assertion failures, we incorporate the relevant code segments extracted during the first stage into the classification process. Specifically, shown as Fig.\ref{Prompt example for assertion classify}, the key to our approach is constructing a multi-modal prompt that integrates three sources of information: the semantics of the assertion, the dynamic behavior from the counterexample waveform, and the logic context provided by RTL code segments. For example, given the assertion \texttt{assert property (@(posedge wb\_clk\_i) (wb\_cyc\_i \&\& wb\_stb\_i) |-> wb\_ack\_o);}, the model can reason about whether the incorrect reason is caused by a timing issue or a logic error by analyzing both the waveform behavior and relevant code. 
\subsection{Stage 3: Incorrect Assertion Fixing}
After identifing the type of errors, the next stage focus on fixing the incorrect assertions. We use tailored prompts designed specifically for each error category to instruct the LLM to generate the correct assertions. This specific approach ensures the fixes precisely address the underlying cause of the assertion failures. Below, we present two specific fix strategies corresponding to the two error types.
\begin{figure}[!t]
\centering
\includegraphics[width=0.7\linewidth]{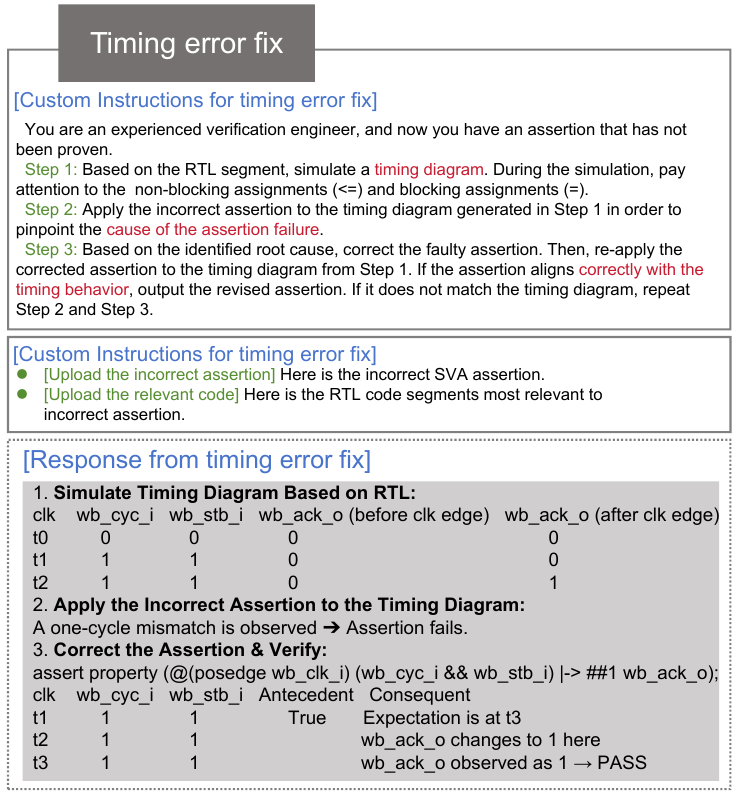}
\caption{Prompt for timing fix}
\label{Prompt example for timing fix}
\end{figure}

\textbf{Timing Error Fix Strategy:} Timing errors typically occur because signals in the assertion are misaligned with their actual timing in the RTL design—for example, signals sampled one cycle too early or late.

To fix such errors, assertions require precise timing alignment with the RTL segments. Therefore, We introduce an automated LLM-based strategy (Fig.\ref{Prompt example for timing fix}) that uses structured prompts to guide the LLM in performing signal-level temporal simulation within RTL segments, identifying timing mismatches between assertions and actual RTL behavior. The prompt instructs the LLM to simulate timing using golden RTL as reference, pinpoint assertion violations, and propose targeted timing adjustments to correct the assertion errors.

\textbf{Logic Error Fix Strategy:} Unlike timing errors that involve signal sampling misalignments, logic errors represent fundamental mismatches between what does the assertion specifies and what the RTL actually implements. These errors occur when assertions miss crucial signal dependencies or encode incorrect logic relationships, resulting in persistent semantic inconsistencies that timing adjustments alone cannot fix.%这里使用encode是因为断言实际上是在用形式语言编码一种行为约束

    Through observation, we find that most fixable logic errors are unidirectional, meaning that either the antecedent or the consequent of the assertion is incorrect. Therefore, to address these fundamental semantic mismatches, we propose a new strategy called \emph{bidirectional anchor reconstruction} (BAR), meaning reconstructing the assertion from both ends (the antecedent and the consequent) simultaneously. Since it might not be clear at first whether the error originates from the antecedent or the consequent, 
\begin{figure}[!t]
\centering
\includegraphics[width=1.1\linewidth]{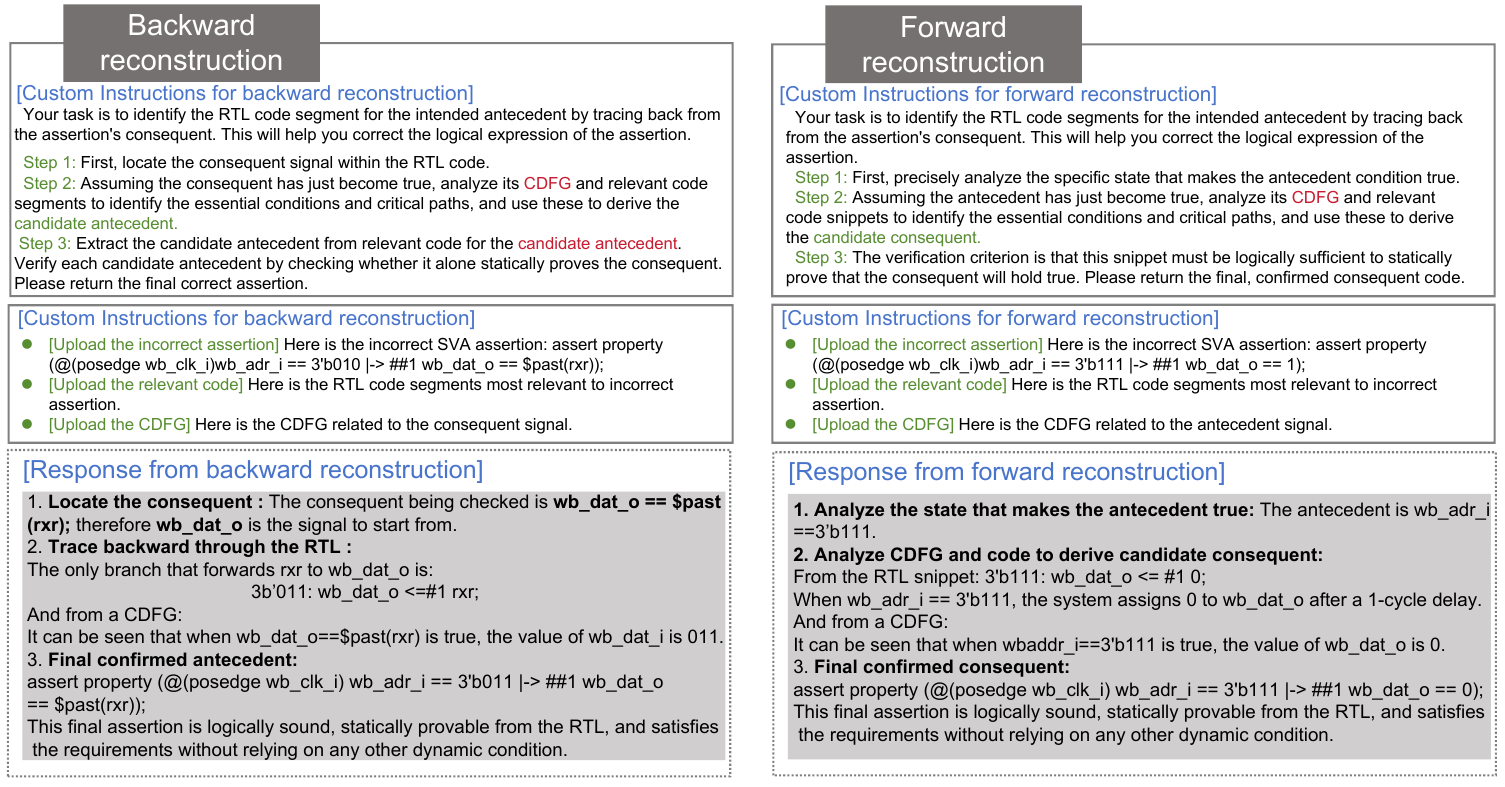}
\caption{Prompt for bidirectional anchor reconstruction}
\label{Prompt example for bidirectional anchor reconstruction}
\end{figure}our approach independently analyzes both directions. The brief process description is as follows:
    \begin{enumerate}
        \item Backward Reconstruction: During backward reconstruction, the consequent of the assertion is assumed to be correct, while the antecedent is considered incorrect or incomplete. A backward analysis is conducted from the signals referenced in the consequent using the CDFG to precisely identify critical logic paths affecting the consequent.
        \item Forward Reconstruction: In forward reconstruction, the antecedent of the assertion is assumed to be correct, and its effects are systematically traced forward through the RTL logic using the CDFG. Analyzing how antecedent conditions propagate to the consequent helps identify deviations caused by incorrectly encoded logic, enabling precise corrections to restore consistency.
    \end{enumerate}
    Fig. \ref{Prompt example for bidirectional anchor reconstruction} illustrates these reconstruction processes using incorrect assertions, demonstrating how backward and forward analyses systematically identify and correct mismatches between the assertion and the RTL segments.
\section{EXPERIMENT}
 In this section,we aim to answer the following research questions (RQ) through a comparative evaluation of AssertFix:
    \begin{enumerate}
        \item \textbf{RQ1:} How does AssertFix perform, in terms of fix correctness, compared with current reasoning-capable LLM?
        \item \textbf{RQ2:} To what extent do the AssertFix improve verification coverage, as measured by COI and proof-core?
        % \item \textbf{RQ3:} Without the \emph{bidirectional anchor reconstruction} strategy, how much does AssertFix’s logic-fix rate decrease?
    \end{enumerate}
\subsection{Experimental setup}
In our study, assertions are initially generated using the state-of-the-art algorithm AssertLLM, which is built upon the gpt-4o-mini-2024-07-18 API. We collect a set of incorrect assertions and make them publicly available via an open-source GitHub repository\footnote{\label{fn:assertfix}\href{https://anonymous.4open.science/r/AssertFix-DB3F/}{Open-source link of AssertFix incorrect assertions}}.%这里的开源描述需要进行一些修改
\setcounter{table}{1}
\begin{table*}[!t]
\centering
\caption{AssertFix and GPT-O3, DeepSeek-R1's rate in assertion fix}%这里需要名字的修改
\label{Number of SVAs: AssertLLM vs. AssertGen}
\begin{tabularx}{\textwidth}{
  >{\centering\arraybackslash}X | 
  *{4}{>{\centering\arraybackslash}X|} 
  *{4}{>{\centering\arraybackslash}X|}
  *{3}{>{\centering\arraybackslash}X|}
  >{\centering\arraybackslash}X 
}
\toprule
\multirow{2}{*}{Metric} & \multicolumn{4}{c|}{\cellcolor{gray!30}\textbf{GPT-O3}} & \multicolumn{4}{c|}{\cellcolor{gray!30}\textbf{DeepSeek-R1}} & \multicolumn{4}{c}{\cellcolor{mycyan!30}\textbf{AssertFix}} \\
\cline{2-13}
& \rule{0pt}{2.5ex}I\textsuperscript{2}C & ECG & \makebox[20pt][r]{Pairing} & \makebox[18pt][r]{SHA3} & I\textsuperscript{2}C & ECG & \makebox[20pt][r]{Pairing} & \makebox[18pt][r]{SHA3} & I\textsuperscript{2}C & ECG & \makebox[20pt][r]{Pairing} & \makebox[18pt][r]{SHA3} \\
\midrule
\makebox[22pt][r]{TE} & 13/8 & 2/0 & 5/1 & 8/3 & 13/9 & 2/1 & 5/3 & 8/3 & \cellcolor{mycyan1!50}13/11 & \cellcolor{mycyan1!50}2/2 & 5/3 & \cellcolor{mycyan1!50}8/7\\
\makebox[22pt][r]{LE} & 24/10 & 17/6 & 22/7 & 22/8 & 24/7 & 17/3 & 22/6 & 22/9 & \cellcolor{mycyan2!30}24/20 & \cellcolor{mycyan2!30}17/12 & \cellcolor{mycyan2!30}22/17 & \cellcolor{mycyan2!30}22/16\\
\makebox[22pt][r]{FR} & 48.6\% & 31.6\% & 29.6\% & 36.7\% & 43.2\% & 21.1\% & 33.3\% & 40.0\% & \cellcolor{mycyan2!30}\textbf{83.8\%} & \cellcolor{mycyan2!30}\textbf{73.7\%} & \cellcolor{mycyan2!30}\textbf{74.1\%} & \cellcolor{mycyan2!30}\textbf{76.7\%}\\
\bottomrule
\end{tabularx}
\end{table*}

To evaluate the correctness of fixed assertions and the coverage improvement brought by the fix, we adopt Cadence JasperGold (version: 21.12.002). Specifically, the formal property verification (FPV) application within JasperGold is used to ensure a comprehensive and unified analysis. All experiments are conducted on a server equipped with an Intel(R) Xeon(R) Gold 6148 CPU @ 2.40GHz.
\subsection{Benchmarks}
To ensure consistency with prior work\cite{b2} and to demonstrate the practicality of our approach, we also select several representative designs from the OpenCores\cite{b18} repository for our experiments, assuming them to be golden RTL. A brief description of each design is provided below, and their corresponding area, cell count, and lines of code (LOC)—synthesized using the open-source tool Yosys-STA\cite{b19} under a 45 nm process node—are summarized in Table I:
    \begin{enumerate}
        \item \textbf{I\textsuperscript{2}C:} A serial communication protocol commonly used for connecting low-speed peripherals.
        \item \textbf{ECG:} A module for biological signal acquisition, typically used in health monitoring systems
        \item \textbf{Pairing:} A cryptographic core for key exchange operations, relevant to secure communications.
        \item \textbf{SHA3:} A hardware implementation of the SHA-3 hash function, used for data integrity and security.
    \end{enumerate}
\setcounter{table}{0}
\begin{table}[h]
\centering
\caption{Benchmarks information}%这里也需要重新起名字
\label{Summary of Design}
\begin{tabular}{|c|c|c|c|}
\hline
\rowcolor{gray!20}
\textbf{Design Name} & \textbf{Area($\mu\text{m}^2$)}& \textbf{Cell number}& \textbf{LOC}\\ \hline
\makecell{\textbf{I\textsuperscript{2}C}} & \makecell{1532.7} & \makecell{756}& \makecell{1282}\\ \hline
\makecell{\textbf{ECG}} & \makecell{89721.3} & \makecell{59084}& \makecell{1635}\\ \hline
\makecell{\textbf{Pairing}} & \makecell{352836.4} & \makecell{228287}& \makecell{2145}\\ \hline
\makecell{\textbf{SHA3}} & \makecell{36346.8} & \makecell{22228}& \makecell{618} \\ \hline
\end{tabular}
\end{table}
\subsection{AssertFix vs. Reasoning LLMs: Fix Rate}
To address \textbf{RQ1}, we analyze the fix rate results of AssertFix compared with two state-of-the-art reasoning-capable LLMs, namely GPT-O3 and DeepSeek-R1, on multiple representative hardware designs. Specifically, we introduce three metrics to comprehensively analyze the assertion-fix performance of different models, namely timing error (TE), logic error (LE), and formal rate (FR). Among these, FR represents the percentage of assertions confirmed as successfully fixed after rigorous formal verification (passing both the "covered" and "checked" stages, i.e., the assertion's antecedent and consequent both pass FPV) using Cadence JasperGold.

Table \ref{Number of SVAs: AssertLLM vs. AssertGen} summarizes the results across four benchmarks. Clearly, AssertFix demonstrates significantly superior performance compared to the two baseline methods, GPT-O3 and DeepSeek-R1, across all four benchmark designs. For instance, in the relatively small-scale I\textsuperscript{2}C design, which includes 13 timing errors and 24 logic errors, AssertFix successfully fixed 11 timing errors and 20 logic errors, achieving an overall FR of up to 83.8\%, whereas GPT-O3 and DeepSeek-R1 only reached accuracies of 48.6\% and 43.2\%, respectively. On the medium-sized SHA3 design, containing 8 timing errors and 22 logic errors, AssertFix also exhibits clear advantages, achieving an FR of 76.7\%, substantially outperforming the best baseline model, DeepSeek-R1 (40\%), further validating the effectiveness and stability of our approach in logic-fix tasks. 

As the design scale and logical complexity increase, the advantage of AssertFix becomes even more prominent. In larger and more complex designs, such as ECG and Pairing (containing a total of 19 errors and 27 errors respectively), AssertFix maintains FR values above 73\%, whereas the fix performance of GPT-O3 and DeepSeek-R1 is significantly limited, with a maximum FR of only around 33.3\%. These results further demonstrate the robustness of AssertFix in handling complex logic fix scenarios.
\subsection{Coverage Gain from AssertFix}
\setcounter{table}{2}
\begin{table*}[t]
\centering
\caption{Comparison of COI/PC between AssertLLM assertion and fixed assertion}%这里需要名字的修改
\label{111}
\begin{tabularx}{\textwidth}{
  >{\centering\arraybackslash}X | 
  *{4}{>{\centering\arraybackslash}X|} 
  *{3}{>{\centering\arraybackslash}X|}
  >{\centering\arraybackslash}X 
}
\toprule
\multirow{2}{*}{Metric} & \multicolumn{4}{c|}{\cellcolor{gray!30}AssertLLM} & \multicolumn{4}{c}{\cellcolor{mycyan!30}AssertLLM with AssertFix} \\
\cline{2-9}
& \rule{0pt}{2.5ex}I\textsuperscript{2}C & ECG & \makebox[20pt][r]{Pairing} & \makebox[18pt][r]{SHA3} & I\textsuperscript{2}C & ECG & \makebox[20pt][r]{Pairing} & \makebox[18pt][r]{SHA3} \\
\midrule
% $FA$ & 84\% & 73\% & 74\% & 76\% & \textbf{133} & 32\% & 41\% & 47\% & \textbf{133} & \textbf{47} & \textbf{103} & \textbf{43}\\
\makebox[22pt][r]{$COI$(\%)} & 86.71 & 99.52 & 33.19 & 89.56 & \cellcolor{mycyan1!50}\textbf{99.11} & \textbf{99.52} & \cellcolor{mycyan1!50}\textbf{99.96} & \cellcolor{mycyan1!50}\textbf{89.61}\\
\makebox[22pt][r]{$PC$(\%)} & 4.43 & 0.22 & 0.05 & 58.90 & \cellcolor{mycyan1!50}\textbf{98.38} & \cellcolor{mycyan1!50}\textbf{16.8} & \cellcolor{mycyan1!50}\textbf{10.2} & \cellcolor{mycyan1!50}\textbf{66.07}\\
\bottomrule
\end{tabularx}
\end{table*}
To answer \textbf{RQ2}, we use two metrics, COI coverage and Proof-Core (PC) coverage, to quantify the improvement in verification coverage after AssertFix. COI coverage ensures the breadth of verification by confirming that all relevant design logic is touched by assertions, thus preventing unexamined blind spots. In contrast, Proof-Core coverage is a stricter metric that measures the subset of logic within the COI that is essential to proving an assertion, thereby ensuring the depth of the verification\cite{b20}.

Table \ref{111} compares the coverage differences between the original assertions generated by the baseline method (AssertLLM, denoted as $COI_{LLM}$ and $PC_{LLM}$) and the assertions fixed by our AssertFix framework (denoted as $COI_{Fix}$ and $PC_{Fix}$). In terms of COI coverage, the largest design, Pairing, shows that 
$COI_{LLM}$ is only 33.19\%, revealing insufficient coverage of key regions; after fix, $COI_{Fix}$ rises to 99.96\%, expanding the coverage by nearly 67 percentage points. For the other three designs (I\textsuperscript{2}C, ECG, and SHA3), $COI_{LLM}$ is already close to or above 85\%, which is a high-coverage zone; AssertFix can still push these values up or keep them around 99\%, showing that even in the “last mile” where small gains are hardest to achieve, AssertFix retains strong fix capability.

Regarding PC coverage, the fix gains are likewise significant and directly reflect the improvement in the logical correctness of the assertions. Taking I\textsuperscript{2}C as an example, $PC_{LLM}$ is only 4.43\%, but after fix it increases to 98.38\%, achieving almost full coverage. This significant improvement is primarily due to the unique structure of the I\textsuperscript{2}C module, whose logic is organized around several key registers (e.g., ctr, sr, rxr, txr, prer, and cr). AssertFix effectively identifies and fixs incorrect assertions related to critical control and data-flow paths associated with these registers, thereby substantially enhancing the overall PC coverage.

Some improvements have also been observed in other designs. For Pairing, $PC_{LLM}$ is merely 0.05\%, whereas $PC_{Fix}$ rises to 10.2\%; ECG likewise climbs from 0.22\% to 16.8\%. For SHA3, the coverage increases from 58.90\% to 66.07\%, a gain of 7.17\% points. Although these values remain low due to inherent limitations in current assertion generation techniques, the improvement clearly indicates that AssertFix effectively reduces the logical errors present in the original assertions, expanding the scope of verified logic and mitigating potential risks of undetected bugs.
\section{conclusion}
This paper addresses the limitation of current LLMs that treat generation as an endpoint and proposes a general-purpose automatic assertion-fixing framework called AssertFix. Built upon LLMs, AssertFix integrates RAG and CoT reasoning to precisely locate RTL segments logically relevant to incorrect assertions, automatically classify error types, and perform semantic fixes on incorrect assertions.

Experimental results demonstrate that AssertFix significantly outperforms SOTA LLMs in terms of assertion fix rate and verification coverage improvement, maintaining high fix effectiveness even as design scale and logical complexity increase.

\end{document}